\documentstyle[aps,epsfig]{revtex}

\begin{document}

\twocolumn

\title{Gravitational lensing of Type Ia supernovae}

\author{
  M. Goliath\thanks{E-mail address: goliath@physto.se}, and 
  E. M\"ortsell\thanks{E-mail address: edvard@physto.se}
}
\address{
  Department of Physics, Stockholm University, \\
  Box 6730, S--113 85 Stockholm, Sweden
}

\maketitle
 
\begin{abstract}
  Recently, Holz and Wald \cite{art:HolzWald1998} have presented a
  method for determining gravitational lensing effects in
  inhomogeneous universes. Their use of realistic galaxy models has
  been limited to the singular, truncated isothermal sphere with a
  fixed mass. In this paper, their method is generalized to allow for
  matter distributions more accurately describing the actual
  properties of galaxies, as derived from observations and N-body
  simulations. This includes the density profile proposed by Navarro,
  Frenk and White, as well as a distribution of galaxy masses. As an
  example of the possible applications of the method, we consider
  lensing effects on supernova luminosity distributions.

  We find that
  results for different mass distributions of smooth dark matter halos
  are very similar, making lensing effects predictable for a broad
  range of halo profiles. We also note, in agreement with other
  investigations, that one should be able to discriminate smooth halos
  from a dominant component of dark matter in compact objects. For
  instance, a sample of 100 supernovae at redshift $z=1$ can, with 99
  \% certainty, discriminate the case where all matter is in compact
  objects from the case where matter is in smooth halos.
\end{abstract}
\vspace{3mm}
PACS numbers: 98.62.Sb, 95.30.Sf, 98.62.Gq, 04.25.Nx

Keywords: Gravitational lensing, Galaxy halos, Supernova observations

\section{Introduction}

Gravitational lensing has become an increasingly important tool in 
astrophysics and cosmology. In particular, the effects of lensing has
to be taken into account when studying sources at high redshifts. 
In an inhomogeneous universe, sources may be magnified or 
demagnified with respect to the case of a homogeneous universe with the same
average energy density.

The effects of gravitational lensing has been studied numerically by a
number of authors, see, e.g.,
\cite{wambsganss,premadi,marri,jain}. The most common method traces
light rays through inhomogeneous matter distributions obtained from
N-body simulations. Lensing effects are accounted for by projecting
matter onto lens planes, and using the thin-lens approximation
(see, e.g., \cite{book:SEF}). 

Recently, Holz and Wald (HW; \cite{art:HolzWald1998}) have proposed
another ray-tracing method for examining lensing effects in
inhomogeneous universes. This method can be summarized as follows: 
First, a Friedmann-Lema\^{\i}tre (FL) background geometry is
selected. Inhomogeneities are accounted for by specifying matter
distributions in cells with energy density equal to that of the
underlying FL model. A light ray is traced backwards to the
desired redshift by being sent through a series of cells, each time
with a randomly selected impact parameter. After each cell, the
FL background is used to update the scale factor
and expansion. By using Monte Carlo techniques to trace a large number
of light rays, and by appropriate weighting \cite{art:HolzWald1998},
statistics for the apparent luminosity of the source is obtained. 

The advantages with this method are that light rays are traced through
a three-dimensional matter distribution without projection onto lens
planes, thus avoiding any assumptions regarding the accuracy of the
thin-lens approximation \cite{kling}.
Furthermore, the method is flexible in the sense that cells may be
taken to represent both galaxies and larger structures with 
different matter distributions, including non-spherical ones. For instance,
HW have performed a number of tests to determine effects of
clustering, and argue that this does not significantly affect statistical
properties of magnification. They also investigate the case of
substructure in the form of compact objects, and conclude that this
can be adequately modelled by randomly distributed compact objects of
arbitrary mass.
It should be pointed out that the method is not well-suited to model 
clustering on scales larger than cell sizes. Still, galaxy clusters can be 
modelled by specifying appropriate masses with corresponding larger cells.
Another drawback is that the method only considers infinitesimal ray
bundles, making it impossible to keep track of multiple images. However, it
is still possible to distinguish between primary images and images that have
gone through one or several caustics \cite{art:HolzWald1998,MGEM-lens}.

HW considered pressure-less models with a cosmological constant, using
the following  
matter distributions: point masses; singular, truncated isothermal 
spheres (SIS); uniform spheres; and uniform cylinders. The individual 
masses were determined from the underlying FL model using a fixed
co-moving cell radius of $R_{c}= 2\mbox{ Mpc}$, reflecting typical
galaxy-galaxy separation length-scales.    

The aim of this paper is to allow for matter distributions 
more accurately describing the actual properties of galaxies.
We will extend the list of matter distributions to
include the density profile proposed by Navarro, Frenk and 
White (NFW; \cite{art:NFW}) and we will use a 
distribution of galaxy masses. 
Also, other matter distribution parameters such as the scale radius
of the NFW halo and the cut-off radius of the SIS halo will be determined
from distributions reflecting real galaxy properties.
The method of HW has also been generalized in Bergstr\"om {\it et al.}
\cite{MGEM-lens} to allow for general perfect fluids
with non-vanishing pressure. 

Gravitational lensing effects may be of importance when, e.g., 
trying to determine cosmological parameters using observations of supernovae
at high redshifts \cite{marri,wambsganssL,metcalf}.
In this paper, we study the effect from lensing on the luminosity 
distribution of a large sample of Type Ia supernovae at redshift
$z=1$.

\section{Mass distribution}\label{sec:massdist}

Realistic modelling of galaxies calls for realistic
mass distributions and number densities, i.e., one has to allow for 
the possibility of the cell radius, $R_{c}$, to reflect the actual distances 
between galaxies. 

An advantage of the method of HW is that it is very easy to allow
for any mass distribution and number density, including possible redshift 
dependencies, as long as the average density agrees with the underlying 
FL-model.
 
Thus, for each cell we obtain a random mass, $M$,
from a galaxy mass distribution $dn/dM$, and calculate the
corresponding radius from the condition that the average energy density
in the cell should be equal to the average matter density of the universe
at the redshift of the cell:
\begin{equation}\label{eq:M}
        M = \frac{4\pi}{3}\Omega_M\rho_{\rm crit}R_{c}^{3},
\end{equation}
where $\Omega_M$ is the normalized matter density, and
$\rho_{\rm crit}=3H^2/8\pi$ is the critical density. 
A galaxy mass distribution can be obtained, for example, by combining
the Schechter luminosity function (see, e.g., Peebles
\cite{book:Peebles}, Eq.~5.129) 
\begin{eqnarray}
  \label{eq:Schechter}
  dn&=&\phi_* y^\alpha e^{-y}dy ,\\
  y&=&\frac{L}{L_*} ,
\end{eqnarray}
with the mass-to-luminosity ratio (see, e.g., Peebles \cite{book:Peebles}, 
Eq.~3.39)
normalized to a ``characteristic'' galaxy with $L=L_*$ and $M=M_*$,
\begin{equation}
  \label{eq:masstolum}
  \frac{M}{M_*}=y^{1/(1-\beta)} .
\end{equation}
Using Eq.~(\ref{eq:Schechter}), we find that
\begin{eqnarray}
  \frac{dn}{dM}&\propto&y^\delta e^{-y} , \\
  \delta&=&\alpha-\frac{\beta}{1-\beta} .
\end{eqnarray} 
Assuming that the entire mass of the universe resides in galaxy halos
we can write 
\begin{equation}
        \int_{y_{\rm min}}^{y_{\rm max}}n(y)M(y)dy = \rho_{m}.
\end{equation}
Using the Schechter luminosity function
and the mass-to-luminosity fraction we get
\begin{equation}
  \label{eq:mref}
        M_* = \frac{\Omega_M\rho_{\rm crit}}{n_{\ast}
          \int_{y_{\rm min}}^{y_{\rm max}}
          y^{\alpha+\frac{1}{1-\beta}}e^{-y}dy}.
\end{equation}
Thus, by supplying values for $n_{\ast}$, reasonably well-determined 
by observations, $y_{\rm min}$ and $ y_{\rm max}$, from which the
dependence of $M_*$ is weak, together with parameters $\alpha$
and $\beta$ we can obtain a $M_*$ consistent with $\Omega_M$.
For the parameter values used in this paper (see Sec.~\ref{sec:results}), 
we get
\begin{equation}
  \label{eq:mrefest}
  M_*\approx 7.5\,\Omega_M\cdot 10^{13}M_{\odot}. 
\end{equation}

\section{The Navarro-Frenk-White distribution}\label{sec:NFW}

In the work of HW, the treatment of 
realistic galaxy models has been limited to the use of the singular,
truncated isothermal sphere (SIS).
Another often-used matter distribution is the
one based on the results of detailed N-body simulations of 
structure formation by Navarro, Frenk and White \cite{art:NFW}.
The NFW density profile is given by 
\begin{equation}
  \label{eq:nfw}
  \rho(r)=\frac{\rho_{\rm crit}\delta_c}
  {(r/R_s)\left[1+(r/R_s)\right]^{2}},
\end{equation}
where $\delta_c$
is a dimensionless density parameter and $R_s$ is a characteristic
radius. The potential for this density profile is given by
\begin{equation}
  \Phi(r)=-4\pi\rho_{\rm crit}\delta_cR_s^2\frac{\ln(1+x)}{x} 
  + {\rm const.},
\end{equation}
where $x=r/R_{s}$. 
The matrix $J^\alpha\!_\beta$, describing the evolution of a light beam
passing through a cell [see Eq.~(37) in HW],
can then be obtained analytically, see \cite{MGEM-lens}.
The mass inside radius $r$ of a NFW halo is given by
\begin{equation}
        M(r) = 4\pi\rho_{\rm crit}\delta_{c}R_{s}^{3}
        \left[\ln (1+x) - \frac{x}{1+x}\right].
\end{equation}
Combining this expression with Eq.~(\ref{eq:M}), i.e., setting $M=M(r)$, 
we obtain
\begin{equation}
        \delta_{c}=\frac{\Omega_M}{3}
        \frac{x_{c}^{3}}{\left[\ln (1+x_{c}) -
        \frac{x_{c}}{1+x_{c}}\right]} 
\end{equation}
where $x_{c}=R_{c}/R_{s}$.
That is, for a given mass $M$, $\delta_{c}$ is a function of $R_{s}$.
From the numerical simulations of NFW we also get a relation between 
$\delta_{c}$ and $R_{s}$. This relation is computed 
numerically by a slight modification of a

\begin{figure}[t]
  \centerline{\hbox{\epsfig{figure=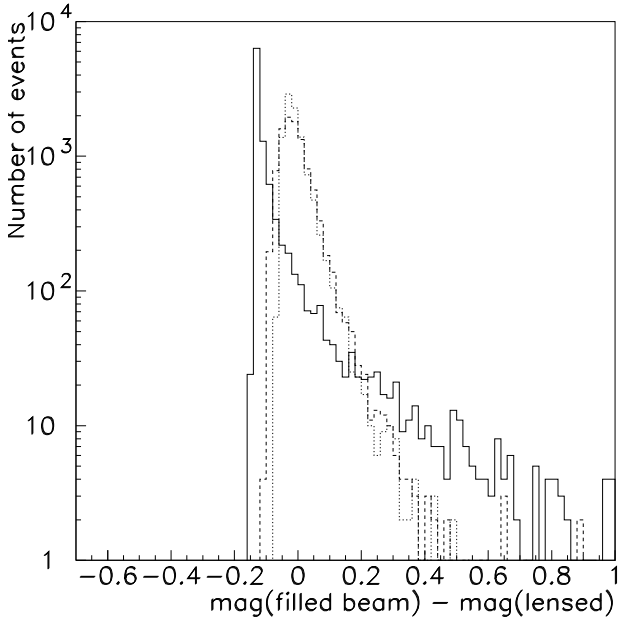,width=0.5\textwidth}}}
  \caption{Luminosity distributions for 10\,000 perfect
    standard candles at
    redshift $z=1$ in a $\Omega_M=0.3$, $\Omega_\Lambda=0.7$
    universe. The magnification zero point is the luminosity in the
    corresponding homogeneous (``filled-beam'') model.
    The full line corresponds to the point-mass case; the
    dashed line is the distribution for SIS halos, and the dotted line
    is the NFW case. This plot can be compared with Fig.~22 of 
    HW.}\label{fig:model-2} 
\end{figure}

\noindent
{\sc Fortran} routine kindly supplied by Julio Navarro. 
Of course, one wants to find a $R_{s}$ compatible with both
the average density in each cell and the numerical simulations of NFW.
Hence, we iteratively determine a value of $R_s$ consistent with both
expressions for $\delta_c$. 
Generally, $R_{s}$ will be a function of mass $M$, the Hubble parameter 
$h$, the density parameters $\Omega_M$, $\Omega_{\Lambda}$, and the
redshift $z$. However, we will use the result 
from Del Popolo \cite{DP99} and Bullock {\em et al.} \cite{BU99} 
that $R_{s}$ is approximately constant with redshift. 
We will compute $R_{s}$ for a variety of $M$, $h$ and 
$\Omega_M$ (all at $z=0$) in both open and flat cosmologies and 
interpolate between 
these values to obtain $R_{s}$ for any combination of parameter values.

\section{Truncation radii for SIS-lenses}\label{sec:SIScutoff}

In their calculations for SIS halos, HW use a fix
truncation radius $d$. However, using a realistic mass distribution,
the cut-off should depend on the mass of the galaxy. Here we derive an
expression for $d$.

The SIS density profile is given by
\begin{equation}
  \rho_{{\rm SIS}}(r)=\frac{\sigma^{2}}{2\pi }\frac{1}{r^{2}},
\end{equation}
where $\sigma$ is the line-of-sight velocity dispersion of the mass particles.
The mass of a SIS halo truncated at radius $d$ is then given by

\begin{figure}
  \centerline{\hbox{\epsfig{figure=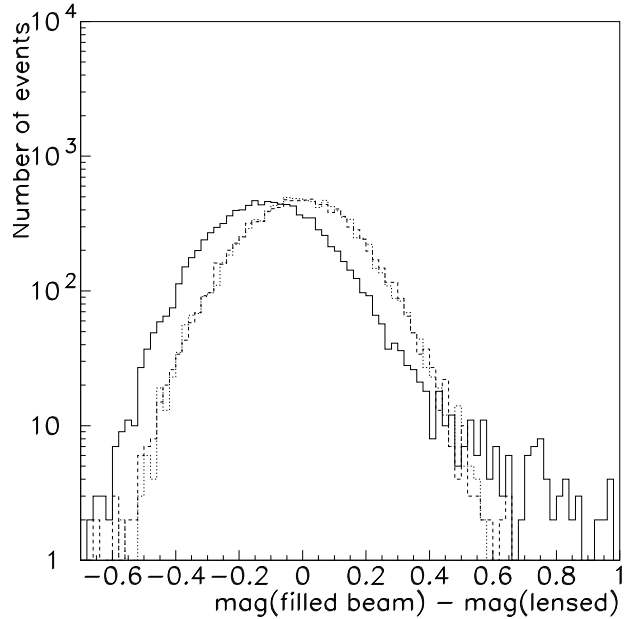,width=0.5\textwidth}}}
  \caption{Luminosity distributions for 10\,000 sources at
    redshift $z=1$ in a $\Omega_M=0.3$, $\Omega_\Lambda=0.7$
    universe. This is the same situation as depicted in
    Fig.~\ref{fig:model-2}, only that we have added an intrinsic 
    luminosity dispersion of the sources with 
    $\sigma_{m} = 0.16$ mag. (corresponding to the case of Type Ia
    supernovae).}\label{fig:intsig-2} 
\end{figure}

\begin{equation}
        M(d) = \int_0^r \rho (r)dV =2 \sigma^{2}d.
\end{equation}
We want this to be equal to the mass $M$ given by the Schechter distribution,
\begin{equation}
        2 \sigma^{2}d=M
        \quad\rightarrow\quad d=\frac{M_*}{2 \sigma_{*}^{2}}
        \left (\frac{M}{M_*}\right )\left (\frac{\sigma}{\sigma_{*}}\right )^{-2},
        \label{eq:d}
\end{equation}
where we, in addition to $M_*$, have introduced a characteristic velocity
dispersion $\sigma_{*}$. Combining the Faber-Jackson relation
\begin{equation}
  \frac{\sigma}{\sigma_{*}}=y^{\lambda}
\end{equation}
with the mass-to-luminosity ratio, Eq.~(\ref{eq:masstolum}), 
we can substitute for $\sigma$ in Eq.~(\ref{eq:d}), and obtain
\begin{equation}
  \label{eq:dcalc}
        d=\frac{M_*}{2 \sigma_{*}^{2}}\; 
        \left (\frac{M}{M_*}\right )^{1-2\lambda (1-\beta)} .
\end{equation}
Using Eq.~(\ref{eq:mrefest}), we can write the truncation radius for
a halo with mass $M=M_*$ as
\begin{equation}
  \label{eq:dest}
  d\approx 3.3\,\Omega_M\, {\rm Mpc}.
\end{equation}

\section{Results}\label{sec:results}

As an application of the method, we investigate lensing effects on
observations of distant supernovae. In Fig.~\ref{fig:model-2}, we
compare the luminosity  
distributions obtained with

\begin{figure}
  \centerline{\hbox{\epsfig{figure=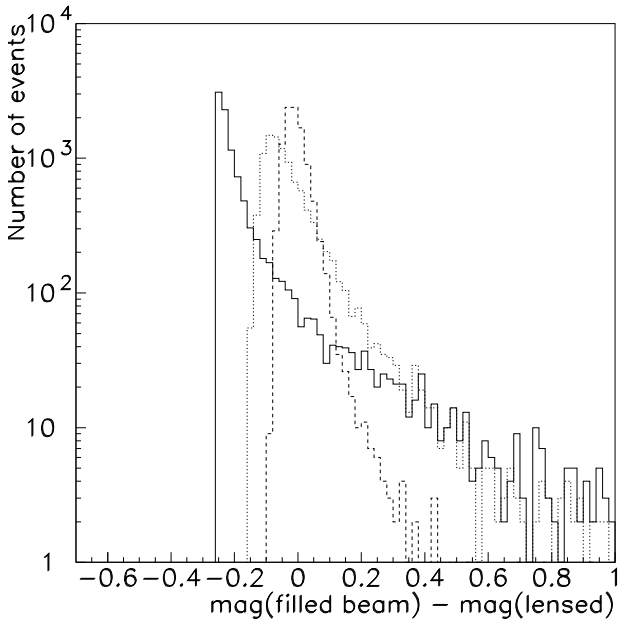,width=0.5\textwidth}}}
  \caption{Luminosity distributions for 10\,000 perfect
    standard candles at
    redshift $z=1$ in a $\Omega_M=1$, $\Omega_\Lambda=0$
    universe. The magnification zero point is the luminosity in the
    corresponding homogeneous (``filled-beam'') model. 
    The full line corresponds to the point-mass case; the
    dashed line is the distribution for SIS halos, and the dotted line
    is the NFW case. This plot can be compared with Figs.~18 and 20 of 
    HW.}\label{fig:model-1}
\end{figure}

\noindent
point masses, SIS lenses and NFW matter 
distributions in a $\Omega_M=0.3$, $\Omega_\Lambda=0.7$ universe,
currently favoured by Type Ia supernova measurements 
\cite{art:Perlmutter-et-al1999,Riess-et-al}.
Sources are assumed to be perfect standard candles.
The magnification, given in magnitudes, has its zero point at the filled beam 
value, i.e., the value one would get in a homogeneous universe. Note that
this value is cosmology dependent. Note also that negative values corresponds
to demagnifications and positive values to magnifications. 
The point mass case (full line) agrees well with the results of Holz and Wald. 
The SIS case (dashed line) is shifted towards the 
filled beam value when comparing to HW. This is due to the fact 
that the cut-off radii computed according to Eq.~(\ref{eq:dcalc}) generally
is much larger than the fix value of $d=200$ kpc used by HW.
Note that results
for SIS halos and NFW halos are very similar, even when we have
no intrinsic luminosity dispersion of the sources.
 
In Fig.~\ref{fig:intsig-2} we have added an intrinsic luminosity
dispersion represented by a Gaussian distribution with $\sigma_m = 0.16$
mag., due to the fact that Type Ia supernovae are not perfect standard
candles. The effect is to make the characteristics of the luminosity
distributions even less pronounced, since the form of the resulting
luminosity distributions predominantly is determined by the form of
the intrinsic luminosity distribution. It is still possible to
observationally distinguish whether lenses consist of compact
objects or smooth galaxy halos, as has been 

\begin{figure}
  \centerline{\hbox{\epsfig{figure=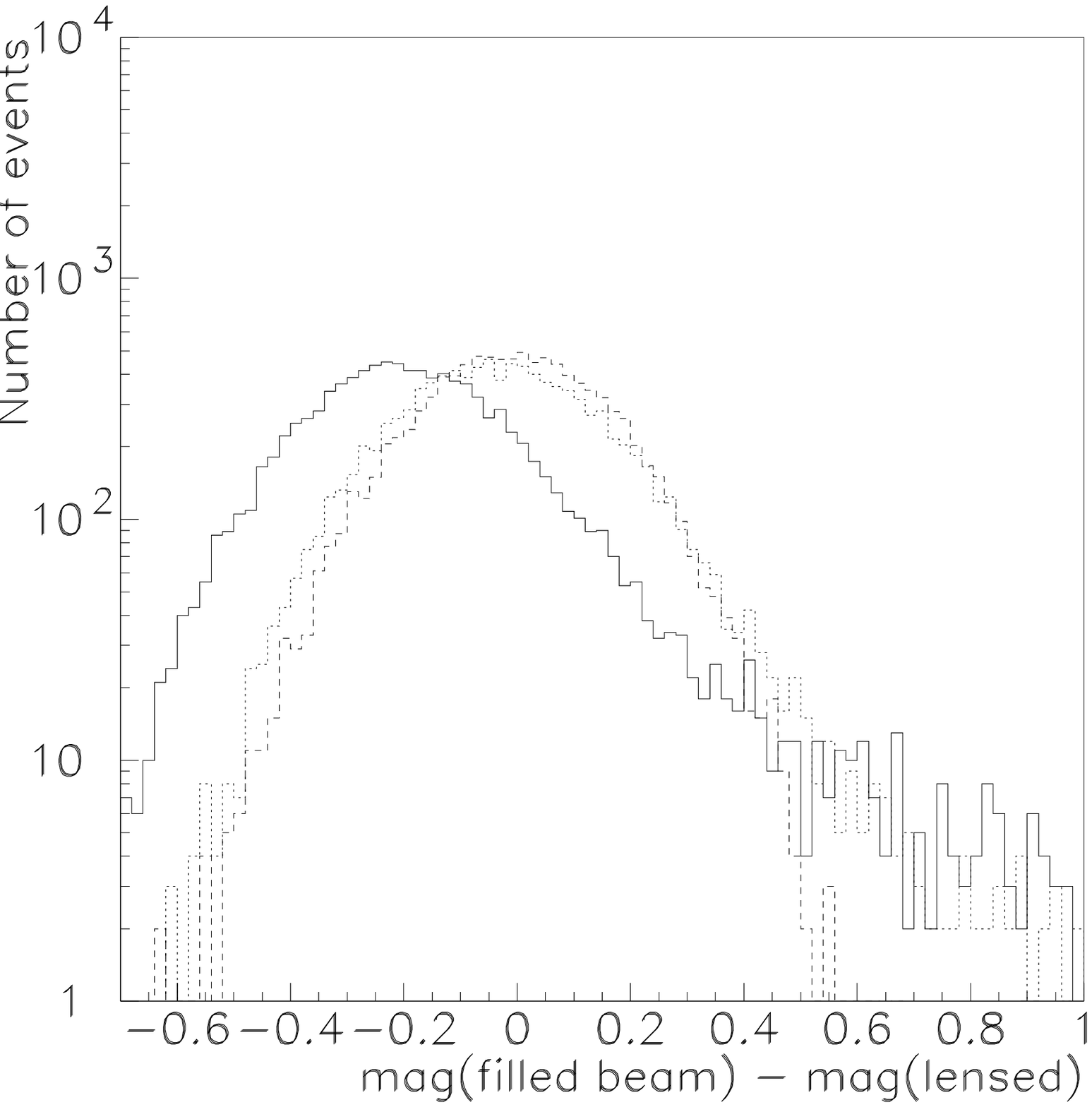,width=0.5\textwidth}}}
  \caption{Luminosity distributions for 10\,000 sources at
    redshift $z=1$ in a $\Omega_M=1$, $\Omega_\Lambda=0$
    universe. This is the same situation as depicted in
    Fig.~\ref{fig:model-1}, only that we have added an intrinsic 
    luminosity dispersion of the sources with 
    $\sigma_m = 0.16$ mag. (corresponding to the case of Type Ia
    supernovae).}\label{fig:intsig-1} 
\end{figure}

\noindent
pointed out in \cite{MetcalfSilk,SeljakHolz}. Generating several samples
containing 100 
supernova events at $z=1$ in a $\Omega_M=0.3$, $\Omega_\Lambda=0.7$
cosmology filled with smooth galaxy halos, we find that for 98 \% of the
samples one can rule out a point-mass distribution with a 99 \%
confidence level\footnote{However, in 1 \% of the samples, we will
  erroneously rule out the halo distribution with the same confidence
  level.}. Furthermore, for a similar sample containing 200
supernovae, the confidence level is increased to 99.99 \%. 

We have performed simulations for various cosmologies, and found a
substantial difference between SIS halos and NFW halos only in a
matter-dominated universe, $\Omega_M=1$, $\Omega_\Lambda=0$, where the
luminosity distribution using NFW halos is shifted towards 
the point-mass case (see Fig.~\ref{fig:model-1}). However, adding an 
intrinsic source dispersion as in Fig.~\ref{fig:intsig-1}, we see that
even with the phenomenal statistics of 10\,000 sources, it would
be a difficult task to
distinguish between the two density profiles.
The increased number of high-magnification events for NFW halos
would probably be the only way to make such a discrimination. 

In these calculations,
we have used the following parameter values (see further \cite{MGEM-lens}):
\begin{itemize}
\item[$\bullet$] $\beta =0.2$
\item[$\bullet$] $\alpha= -0.7$
\item[$\bullet$] $y_{{\rm min}}=0.5$
\item[$\bullet$] $y_{{\rm max}}=2.0$
\item[$\bullet$] $n_*=1.9\cdot 10^{-2}\, h^{3}{\rm \ Mpc}^{-3}$
\item[$\bullet$] $\sigma_{*}=220\mbox{ km/s}$ 
\item[$\bullet$] $\lambda =0.25$
\end{itemize}

A more extensive discussion of the luminosity distributions of perfect
standard candles obtained 
with the different halo models at different source redshifts can be 
found in Bergstr\"om {\em et al.} \cite{MGEM-lens}, where also some analytical
fitting formulas for the probability distributions are given.

\section{Discussion}\label{sec:disc}

In this paper, the method of Holz and Wald \cite{art:HolzWald1998} has 
been generalized to allow
for matter distributions reflecting the actual properties of 
galaxies, including the density profile proposed by Navarro, Frenk and 
White \cite{art:NFW}. 
In order to make matter distributions as realistic
as possible, all parameter values in the lens models are obtained from 
reasonable probability distributions, as derived from observations and
N-body simulations. This includes the mass of the galaxies,
the truncation radius of SIS lenses and the characteristic radius of NFW
halos.
One of the virtues of this method is that it can be continuously refined
as one gains more information about the matter distribution in the 
universe from observations.

The motivation for these generalizations
is to use this method as part of a model for simulation of
high-redshift supernova observations.
In this paper, we have considered lensing effects on supernova luminosity 
distributions. 
Results for different mass distributions in smooth dark matter halos
was found to be very similar, making lensing effects predictable for
a broad range of density profiles.
Furthermore, given a sample of 100 supernovae at $z\sim 1$, one should
be able to discriminate between the case with smooth dark matter halos
and the (unlikely) case of 
having a dominant component of dark matter in point-like objects.

\section*{Acknowledgements}

The authors would like to thank Daniel Holz and Bob Wald for helpful
comments, and Julio Navarro for providing his numerical code
relating the parameters of the NFW model.

\end{document}